\documentstyle[prl,aps,multicol,rotate,epsf]{revtex}
\begin{document}
\draft
\title{ Phase Transitions of Soluble Surfactants at a Liquid-Vapor Interface}
\author{M. S. Tomassone$^1$, A. Couzis$^2$, C. Maldarelli$^{1,2}$, J. R. 
Banavar$^4$ and J. Koplik$^{1,3}$.}
\address{Benjamin Levich Institute$^1$ and Departments of Chemical
Engineering$^2$ and Physics$^3$ \\
City College of the City University of New York, New York, New York, 10031\\
$^4$Department of Physics and Center for Materials Physics\\
The Pennsylvania State University, University Park, PA 16802}
\date{\today }
\maketitle

\begin{abstract}
Although medium chain length insoluble amphiphiles are well known to 
form gaseous and liquid expanded phases on an air/water interface, the 
situation for the soluble case is less clear. 
We perform molecular dynamics simulations 
of model surfactant molecules dissolved in a bulk liquid solvent 
in coexistence with its vapor.  Our results indicate a transition in 
both soluble and insoluble surfactants:  a plateau in surface tension vs.\
surface coverage, whose instantaneous configurations display two phase
coexistence, along with correlation functions indicating a transition to 
gaseous to liquid-like behavior. 
\end{abstract}

\begin{multicols}{2}
\narrowtext
Surfactants are amphiphilic molecules which contain a polar 
head group and a nonpolar tail, which usually consists of 
a chain of hydrocarbon groups.  At an air/water interface
surfactants arrange themselves
with their polar group immersed in water, interacting by dipolar forces, 
while the hydrocarbon tails are displaced outward into the air.
The presence of surfactants at the air/water interface lowers the
free energy, and in turn the surface tension, usually as a non-increasing
function of the concentration of surfactant in the liquid \cite{adamson}.
The degree of aggregation and molecular ordering of surfactants adsorbed 
at an air/solvent interface is reflected in the appearance of distinct
surface phases as a function of surfactant concentration.
In addition to the obvious two dimensional Gas (G) and disordered liquid 
phases (more precisely, the Liquid Expanded LE phase), it is 
possible to find condensed mesophases and semi-solid crystalline-like
phases in monolayers \cite{knobler,kaganer}.

This paper focuses on the determination of {\it soluble} surfactant 
adsorption isotherms
simultaneously with their phase behavior and surface tension variation,
and in particular the G/LE transition, using Molecular 
Dynamics (MD) simulations.  In the context of 
insoluble surfactants, this approach has been used extensively to study
the structure of monolayers restricted to the surface of a substrate 
\cite{klein,karaborni,rice},
obtaining information both complementary to and in semi-quantitative 
agreement with experiment on Langmuir films.  For soluble surfactants 
there is emerging experimental evidence for the existence of gaseous,
liquid, and condensed surface phase transitions \cite{soluble},
but no accompanying theory or simulation.  There are
MD simulations of soluble surfactant monolayers \cite{bocker}, but these do
not study the surface-bulk equilibrium which gives rise 
to the surface phases, or the transitions between them.
In contrast, MD simulations in liquid-liquid systems \cite{smit} have studied
the equilibrium between the liquid phase and monomeric and surfactant
aggregates in the bulk, but not surface phase transitions.

Our simulations model surfactant molecules placed 
in a solvent in equilibrium with its vapor,
which then migrate to the surface to a degree controlled by the choice of
interaction potential.
We observe that as their initial concentration increases, the 
surfactant structure on the liquid/vapor interface ranges from
a gas phase containing isolated molecules or small clusters, to a coexisting
mixture of clusters of various sizes, to a single disordered spanning liquid
cluster.  Examination of positional and tail-orientation distribution functions
show a gradual transition to liquid-like ordering, and the isotherm of
surface tension {\em vs}. surface concentration exhibits a plateau 
indicative of a phase transition.  We have studied both insoluble and 
soluble surfactant systems, obtaining rather similar behavior, but in 
this Letter we concentrate on the latter.  Further details will be presented 
elsewhere \cite{long}. 

\begin{figure}
\centerline{
\hbox{
\epsfxsize=7.5cm \epsfbox{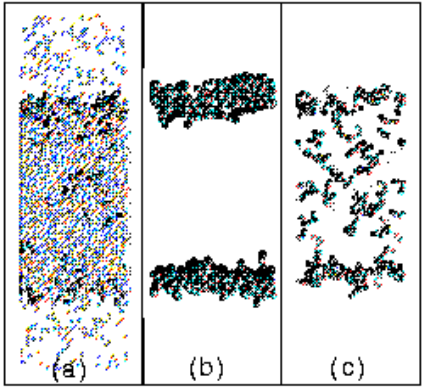}}} 
\vspace{.5cm}
\caption{Snapshots of the simulated system;  (a) all atoms in a soluble case, 
(b) surfactant molecules only in an insoluble case, (c) surfactant molecules
only in a soluble case.}
\label{full}
\end{figure}

Our calculations are based on standard MD techniques
using Lennard-Jones (LJ) interactions, in an NVT ensemble \cite{at}.
The potential between any two atoms of type $i$ and $j$ 
separated by distance $r$ is 
$V^{LJ}_{ij}(r) = 4 \varepsilon [ \left(\sigma/r \right)^{12}
-C_{ij} \left( \sigma/r \right)^6 ]$. Here 
$\epsilon$ is an energy scale, $\sigma$ is approximately an atomic diameter,
and the characteristic microscopic time unit is 
$\tau=\sqrt{(m\sigma ^2/\varepsilon )}$, with atomic mass $m$. 
The adjustable coefficients $C_{ij}$ 
determine the interactions between the various molecular species and 
control the chemistry of the system.  The surfactant atoms are bound in chain 
molecules using the Finitely Extensible 
Nonlinear Elastic (FENE) potential \cite{fene} 
$V^{FENE}=-\frac{1}{2}K r_0^2\; \ln{\left[ 1- (r/r_{0})^{2}\right]}$
which acts only between adjoining atoms on a chain;  
$r_0$ is the maximum bond length and $K$ is a spring constant.
The system is maintained at a temperature $T=0.9\epsilon/k$ by a 
constant kinetic energy thermostat.  
We have worked with a system of 11520 atoms in total, with surfactant  
chains of length 6.  The atoms move in a three dimensional box of
size $20.5\,\sigma \times$ $20.5\,\sigma \times 68.4\,\sigma$, with 
periodic boundary conditions in all directions, and form a horizontal liquid
slab with vapor above and below, with two (statistically) planar liquid/vapor
interfaces.  A snapshot of
a typical configuration is shown in Fig.~\ref{full}a.

\begin{figure}
\centerline{ 
\hbox{
\epsfxsize=5.cm \rotate[r]{\epsfbox{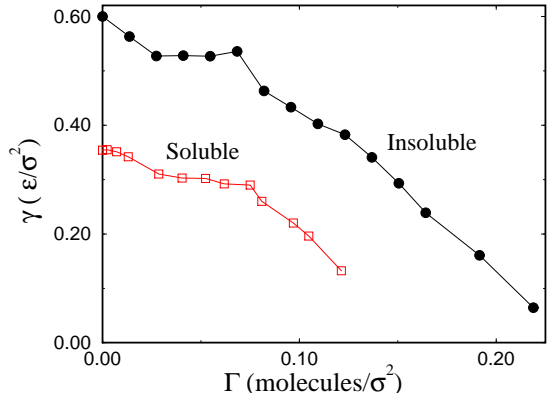}}}} 
\vspace{.4cm}
\caption{$\gamma -\Gamma$ isotherms for the insoluble and soluble cases.}
\label{isotherm}
\end{figure}

The surfactant molecules are initially distributed at random in the liquid,
and then
migrate to the liquid-vapor interface to a degree determined by the choice
of $C_{ij}$, so that the solubility is indirectly under control.
An {\em in}soluble surfactant is
achieved by choosing a strong attraction between head group and solvent, 
an antipathy between surfactant tail and solvent, and a strong cohesion
among the molecules in the solvent.  The full set of interaction
coefficients is shown in Table~I, where the subscript convention is {\bf 1}: 
hydrophilic head, {\bf 2}: hydrophobic tail and {\bf 3}: solvent.   
A typical snapshot of the insoluble surfactant phase alone is shown in 
Fig.~\ref{full}b.
For a soluble surfactant system, we modify the interactions 
to allow surfactant to remain in solution. 
A first ingredient is to reduce the antipathy between the surfactant tails 
and the solvent, by increasing $C_{23}$, and the second is to
reduce the cohesion of the solvent, by reducing $C_{33}$, and the full set
is shown in Table~II, 

The surface tension is computed from the standard microscopic expression 
for a planar interface \cite{row-widom},
$\gamma= \frac{1}{A}\left< \sum_{i<j}
\frac{r_{ij}^{2} - 3 z_{ij}^{2}}{2 r_{ij}} \frac{dV(r_{ij})}{dr_{ij}}
\right>$,
where the sum runs over all interacting pairs of atoms, $V$ is the 
full interatomic potential, $A$ is area of the interface,
and the angle brackets denote a time average.  
The insoluble surface tension isotherm is given in Fig.~\ref{isotherm}.
There is a clear reduction for small amounts of surfactant,  
followed by an approximate plateau until a critical concentration 
is reached, and then a rapid decrease.  (At still higher surfactant
concentration we observe buckling of the interface or micelle formation.)
The structure of the surfactant layer may be inferred from the pair
distribution function $g(r)$ \cite{hd}, here taken to be the probability 
that two head groups on a given interface are separated 

\medskip

\begin{figure}
\centerline{
\hbox{
\epsfxsize=5.cm \rotate[r]{\epsfbox{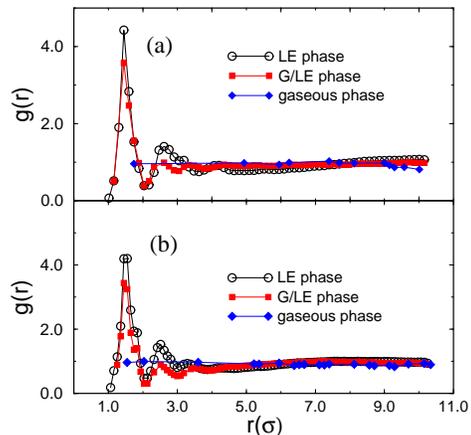}}}} 
\vspace{.7cm}
\caption{Pair distribution function for surfactant monolayers on the 
liquid-vapor interface: (a) insoluble, (b) soluble.}
\label{grsoluble}
\end{figure}

\begin{tabular}{||r|rrr||r||r|rrr||r||}
\multicolumn{5}{c}{Table I: Insoluble.}\hspace{1cm}&
\multicolumn{5}{c}{Table II: Soluble.}\\ \cline{1-4} \cline{6-9}
C(i,j)&{\bf 1}&{\bf 2}&{\bf 3}&&C(ij)&{\bf 1}&{\bf 2}&{\bf 3}\\ 
\cline{1-4} \cline{6-9}
{\bf 1} &  1.0  &1.0  &3.0    &&{\bf 1} & 1.0&1.0&3.0\\ 
{\bf 2} &  1.0  &0.2  &0.6    &&{\bf 2} & 1.0&0.2&0.8\\
{\bf 3} &  3.0  &0.6  &1.15   &&{\bf 3} & 3.0&0.8&1.10\\ 
\cline{1-4} \cline{6-9}
\cline{1-4} \cline{6-9}
\end{tabular}
by a distance $r$. As we see in Fig.~\ref{grsoluble}a, in the very low coverage
region $g(r)$ is a flat line typical of a gas, but as the coverage increases
up to the end of the plateau, the curve changes to that of a typical
liquid, with a prominent nearest-neighbor first peak, a rather broad second 
peak exemplifying short range order, and some further structure at larger
$r$ as well.  This simulation therefore reproduces the
well-documented G/LE phase transition for insoluble surfactant.

\begin{figure}
\centerline{
\vbox{ \hbox{\epsfxsize=4.5cm {\rotate[r]{\epsfbox{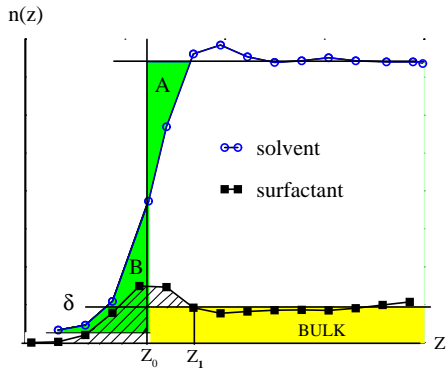}}}}
       }
}
\vspace{.3 cm}
\caption{Interface profile obtained in the simulation, and the Gibbs 
construction.}
\label{scheme}
\end{figure}

For the soluble surfactant system, a representative snapshot  
of the positions of surfactant molecules is in Fig.~\ref{full}c. The
measured isotherm is shown in 
Fig.~\ref{isotherm}, and has the same qualitative features as in the
insoluble case.  
In this figure, the surfactant surface coverage is determined by considering 
the component 
density profiles as a function of distance normal to the interface, 
Fig.~\ref{scheme}.  The upper curve is the solvent profile, while the lower
one refers to the tail groups, and the surface coverage $\Gamma$ is
taken to be the total area under the bump to the left of $z_1$.  
To obtain the surfactant adsorption isotherm in a 
thermodynamically unambiguous manner, we use the density profiles
to extract Gibbs' ``surface excess concentration'' $\Gamma_e$ \cite{adamson}. 
We first locate a Gibbs dividing surface, $z_{0}$,
which serves as a convenient definition of the solvent interface. 
Referring to Fig.~\ref{scheme},
the location of this surface is defined by the condition that the areas
of the shaded regions $A$ and $B$ are equal, thereby requiring the surface
excess of solvent to vanish.   The bulk surfactant
concentration is defined by extrapolating the average bulk density $\delta$
in the bulk up to the Gibbs dividing surfaces, 
\begin{figure}
\centerline{
\hbox{
\epsfxsize=4.7cm \rotate[r]{\epsfbox{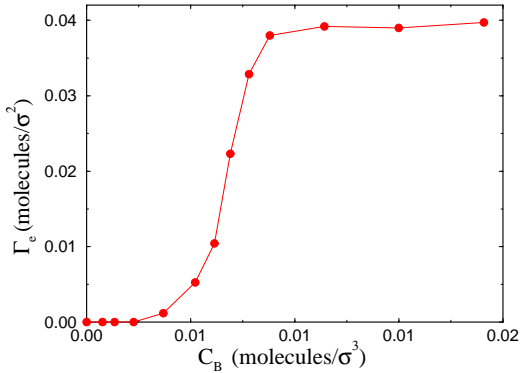}}}}
\vspace{.5cm}
\caption{Surface concentration vs. Bulk concentration.}
\label{sigmoid}
\end{figure}
and the surface excess concentration of surfactant $\Gamma_e$ 
is then the remainder, -- the area of the hatched region in the figure.
(We assume that there is no surfactant in the vapor phase.) 
Repeating this procedure at each value of bulk concentration, we obtain
the variation of surface excess coverage with total concentration shown in 
Fig.~\ref{sigmoid}: a monotonic increase, with a
sigmoidal shape.  Note that the steeply rising part of the curve occurs
precisely in the plateau region of the isotherm, which we will argue 
corresponds to the G/LE phase transition.  In the thermodynamic limit
one would expect a vertical line there, as surfactant is expelled from
solution to form larger liquid-like clusters on the interface,  
but we observe the smoothing and finite slope
characteristic of phase transitions in finite-sized systems.

\begin{figure}
\centerline{
\hbox{
\epsfxsize=9.cm \epsfbox{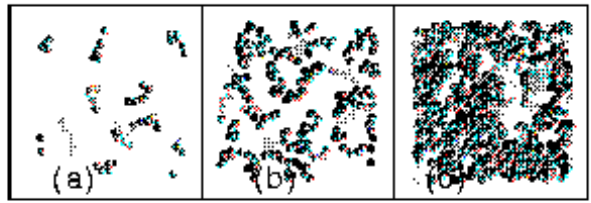}}} 
\vspace{.5cm}
\caption{Snapshots of soluble surfactant molecules at an interface:
(a) gas,(b) gas-liquid and (c) liquid expanded phases.}
\label{solubles1}
\end{figure}

Returning to the soluble isotherm, 
note that the critical densities of the gas and liquid phases for the G/LE
transition (the end points of the plateau) occur near the same values
of $\Gamma$ found for the insoluble case, presumably because this value 
corresponds to the onset of clustering, and is determined largely by the 
attraction between surfactant molecules, which was not changed.
The pair distribution function for systems at the start, middle and end
of the surface tension plateau are shown in Fig.~\ref{isotherm}b, and
generally resemble the insoluble case.  The characterization of these phases
is facilitated by examining snapshots of the molecular configuration, and
three typical examples corresponding to the three surface coverage values 
where $g(r)$ was plotted are shown in Fig.~\ref{solubles1}.  
At low surface concentration, Fig.~\ref{solubles1}a at $\Gamma=0.014$, one sees 
a gaseous state with isolated molecules and a weak tendency to 
bind into pairs.  At higher coverage $\Gamma=0.062$ in the middle of 
the plateau, Fig.~\ref{solubles1}b, 
we see a spanning cluster of surfactant molecules, accompanied by isolated
molecules. Finally at the end of the plateau, $\Gamma=0.097$ in  
Fig.~\ref{solubles1}c, there is a single surfactant monolayer.

\begin{figure}
\centerline{
\hbox{
\epsfxsize=5.cm \rotate[r]{\epsfbox{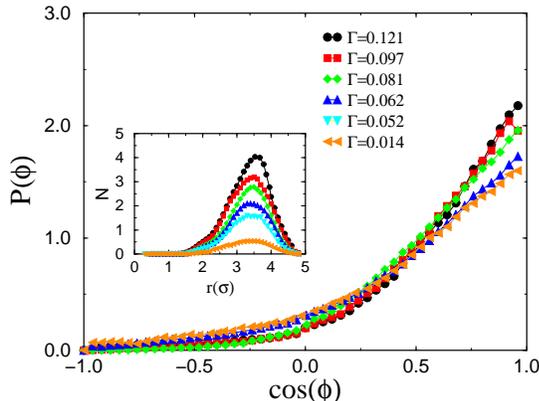}}}}
\vspace{.5cm}
\caption{Probability distributions of the tilt angles, at various values of
surface concentration.  The inset shows the corresponding
histogram of the length of the molecules' end-to-end vectors.}
\label{tilt}
\end{figure}

Further evidence for phase transitional behavior in the plateau region  
is obtained by studying the
degree of ordering of the surfactant tails. A simple quantitative
measurement is the histogram of the tilt angle between the surfactant tail 
and the normal to the interface, shown in Fig.~\ref{tilt}.
The orientation of a molecule is defined as the direction of the eigenvector
of its moment of inertia tensor with the smallest eigenvalue \cite{karaborni}. 
We see that as the surface coverage increases across the plateau, molecules 
are increasingly likely
to be oriented normal to the interface.  An even more dramatic variation
is seen in the histogram of end-to-end lengths, shown in the inset to 
Fig.~\ref{tilt} -- as coverage increases, the molecules are increasingly
extended as they pack together.  Note that because the molecular
interaction corresponds to freely-jointed chains with no bond-bending
forces, the surfactant molecules need not form straight rods,
and we do not observe complete correlation of the tails seen in MD
simulations of condensed phases in Langmuir monolayers \cite{karaborni}.  
Furthermore, since the tails are now partially soluble they are more 
likely to have  solvent around them, which tends to decorrelate 
their orientation.
\begin{figure}
\centerline{ 
\hbox{
\epsfxsize=5.cm \rotate[r]{\epsfbox{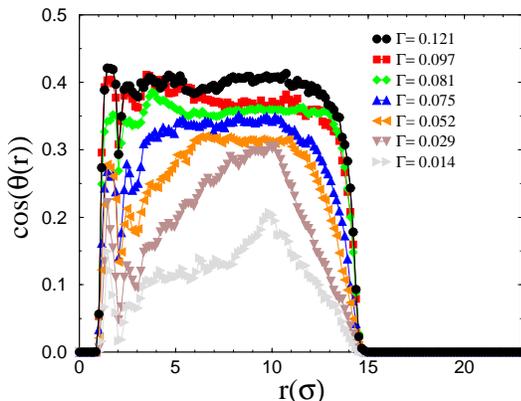} }} }
\vspace{.5cm}
\caption{Orientational correlation function at various coverages.}
\label{compacorre}
\end{figure}

Lastly, we can go further in this vein and define a pair angular correlation
function: the probability distribution for the mean 
relative angle between two molecular tails as a function of the head group
separation $r$.  Results are shown in Fig. \ref{compacorre}:
at low coverage, there is little correlation
between the tails, but the likelihood of alignment increases with
$\Gamma$, as the molecules pack together and the tail are attracted to 
each  other.

In summary, we have presented MD simulations of soluble amphiphilic 
systems, containing simple chain molecules in a monatomic bulk solvent,
which assemble from solution into a surfactant layer at the liquid-vapor
interface.  As a function of surface coverage, these systems exhibit a 
G/LE phase transition which can be characterized by plateaus in the surface
tension isotherm, visible changes in the molecular configurations, and 
corresponding transitions in various correlation functions.  
The behavior of insoluble and soluble systems is found to be very similar,  
suggesting that at least this surface ordering phenomenon is not highly 
sensitive to the dynamics in the bulk solution, but rather an effect of
surface packing.  

\medskip

We thank H. A. Makse for discussions, S. McNamara for his collaboration in
the early stages of this work, NASA and NSF for financial support, and NPACI
for computer resources.

\end{multicols}

\end{document}